\documentclass{aa}

 \usepackage{psfig}
\usepackage{aabib99}
\begin{document}
\thesaurus{02 (12.04.2;    
               11.05.2;    
               11.06.1)    
          }

\title{Cosmological history of stars and metals}
\author{R. Sadat\inst{1}, B. Guiderdoni\inst{2} and J. Silk\inst{3} } 

\institute{ Observatoire de Midi-Pyr\'en\'ees, 14 avenue Edouard Belin, 31400 - Toulouse, France 
          \and 
           Institut d'Astrophysique de Paris, 98bis Boulevard Arago, F-75014 Paris, France
         \and
	Department of Physics, Astrophysics, 1 Keble Road, Oxford OX1 3NP, 
England }

\offprints{R. Sadat}
\date{Received   \rule{2.0cm}{0.01cm} ; accepted \rule{2.0cm}{0.01cm }}

\maketitle

\begin{abstract}
	 We study the evolution of stellar content and the chemical enrichment of the universe averaged over the whole population of galaxies by means of a series of chemo--spectro\-photometric models that take into account the metallicity and dust obscuration effects. We investigate various classes of cosmic star formation rates (CSFR) histories consistent with current estimations. We are able to reproduce a variety of observational constraints such as the emissivities at rest--frame 0.44, 25, 60 and 100 $\mu$m of the local universe and also the overall shape of the extragalactic background light from UV/NIR galaxy counts and the cosmic infrared background (CIB) from DIRBE/FIRAS measurements. We find that the CIB at 140 $\mu$m is crucial for discriminating between the CSFR histories. The best--fit model to this constraint seems to favour mid--infrared derived CSFR at low--$z$ and a flat CSFR at higher $z$ consistent with the most recent estimations, although the shape of the CSFR at high redshifts has little impact on the FIR/submm part of extragalactic background. We suggest that the bulk of the CIB energy is produced by a population of moderately obscured normal galaxies lying at $0\le z \le 1$. We then derive the global chemical enrichment and stellar content of the universe and find that this model predicts metallicities in good agreement with the metallicity in DLAs given that some outflow of metal--enriched gas from galaxies is assumed, but it overproduces the current present--day stellar mass density and NIR luminosity density.  
\end{abstract}


\maketitle
\section{Introduction}
\hspace{0.5cm}
	The history of the cosmic star formation rate (CSFR) in the universe has been inferred in an unprecedented 
way by combining the SFR in the local universe, the results of ground--based spectroscopic 
redshift surveys out to $z \sim 1$ (Lilly {\it et al.} 1996) and the results of deep photometric 
surveys like the {\it Hubble Deep Field} (HDF) in which galaxies photometric redshift have been 
determined by the Lyman dropout selection technique (Madau {\it et al.} 1996, hereafter M96). The 
resulting cosmic star formation history has been summarized in the so--called ``Madau 
plot'' which shows a rapid increase up to $z\sim1$ and then a decrease at higher $z$ with a peak
at $z \sim $1.5 - 2. However, one of the large uncertainties in deriving the CSFR from the 
comoving luminosity density is dust obscuration. Indeed, the faint objects have been selected optically and therefore may represent only the tip of the iceberg because dust absorbs large 
amounts of UV--optical rest--frame light, causing underestimation of the SFR by 
a large factor. This factor is still a matter of debate, for exemple Pettini {\it et al.} (1998) have estimated the extinction on the Lyman Dropout galaxies (at $z >2$) and found 
a factor of  $\sim 3$ while Meurer {\it et al.} (1997) derived for the same objects a factor as large 
as 15. Finally, there might exist a population of star--forming galaxies at high--$z$ which are completely enshrouded in dust and therefore remain undetected in the optical/near infrared as revealed by the ISOCAM--HDF survey (Aussel {\it et al.} 1998, Elbaz {\it et al.} 1999 ). Moreover, new estimates of the the global star formation rate have been reported. The first one is derived from the analysis of deep ISO images of the CFRS by Flores {\it et al.} (1999, hereafter F99) who revised previous star formation density by a factor of $\sim 3$ upward at $0 \le z \le 1$ and the second one comes from an analysis of Lyman Break Galaxies in the HDF by Steidel {\it et al.} (1999, hereafter S99) who found a higher SFR at $z \approx 4$ than previous estimate by Madau {\it et al.} 1998 (hereafter M98) and therefore found no evidence for a significant decline of the CSFR at high redshifts.\\
The history of CSFR can also be observed by its imprint on the background radiation. The first 
detection in DIRBE data of the IR/submm background by Puget and his collaborators (1996) at a 
level comparable and even higher than the optical counterpart, suggests that a large fraction of 
the energy of young stars is absorbed by dust and reradiated in the IR/submm. Understanding the 
nature and redshift of the sources responsible for the infrared background is of great 
importance for galaxy evolution studies. Several deep surveys have been undertaken in the submm 
wavelengths range at 450 and 850 $\mu$m with the SCUBA Bolometer Array and ISO satellite (at 175 
$\mu$m), in order to resolve this background (Kawara {\it et al.} 1997, Barger {\it et al.} 
1998, Hughes {\it et al.} 1998, Eales {\it et al.} 1999).\\
Recent advances in high resolution spectroscopy of high redshift systems 
like damped Ly${\alpha}$ systems (DLAs), has provided us with precious information on early phases of 
chemical enrichment (Pettini {\it et al.} 1997, hereafter P97) and neutral gas content of the 
universe (Storrie-Lombardi {\it et al.} 1996) casting some light on the cosmic chemical 
evolution. 
The aim of this paper is to explore the predictions that models of stellar evolution self--consistently coupled with chemical evolution code would make on the metal and stellar enrichment history of the universe. Briefly, given the cosmic star formation history we calculate the unattenuated spectral luminosity density at different redshifts. Adopting a model of dust absorption we then compute the attenuated comoving luminosity density at different redshifts and wavelengths which is then compared to available observations. Combining this with the extragalactic background will allow us to constrain the models. Predictions on the cosmic chemical enrichment and on the stellar content of the universe are then derived. \\
This paper is organized as follow: section 2 describes the models and the inputs we used. In section 3, we discuss our models predictions in the light of available photometric constraints, such as the comoving luminosity densities from UV/vis to FIR and the cosmic infrared background radiation (CIB). In section 4 we derive the chemical enrichment history of the universe and its stellar content. Throughout the paper we use a value 
of the Hubble constant H$_{0}=50$ km s$^{-1}$ Mpc$^{-1}$, a density parameter $\Omega_{0}=1$ and 
$\Omega_\Lambda=0.$ 
\section{The models}
\hspace{0.5cm}
In this work we study the global properties of galaxies averaged in large comoving volumes. We are in particular interested in the stellar and heavy element content of the universe that our models would predict using available observational constraints. This is based on chemo--spectrophotometric models that follow in a self--consistent way the evolution of stars, dust and metals. These models are based on stellar population synthesis code which is coupled to simple chemical evolution models and dust absorption. The basic ingredients of our models are described bellow (more details can be found in Devriendt {\it et al.} 2000, hereafter DGS)   
\subsection{The stellar emission}
\hspace{0.5cm}
	Our models are based on up--dated libraries of theoretical stellar tracks and spectra allowing for the building of new sets of spectrophotometric evolutionary models that take into account the 
 effect of metallicity. These models use the Geneva group stellar evolutionary tracks 
 (Schaller {\it et al.} 1992, Schaerer {\it et al.} 1993a,b) for five metallicities 
$Z =0.001$, 0.004, 0.008, 0.02, 0.04. The main stellar evolution phases include all the 
important stages except for low mass stars (less than 1.7 M$_{\odot}$) those tracks stop at 
the Giant Branch tip. We thus use the more recent track evolution of low mass stars which 
go from the Zero--main sequence up to the end of EAGB for only two metallicities $Z=0.02$ 
and $Z=0.001$ (Charbonnel {\it et al.} 1996). For the other metallicities we simply 
interpolate between the available Geneva tracks. Choosing the Geneva group tracks allows the use of the $Z$--dependent stellar yields of Maeder (1992) which insures the consistency when coupling spectrophotometric and chemical evolutions. For the stellar spectra we use theoretical spectra by Kurucz (1992), with the advantage that it avoids the use of
the transformations of bolometric luminosity log L$_{bol}$ to magnitude 
M$_{V}$, of log T$_{eff}$ to Spectral Type, and to luminosity classes, which are rather uncertain for the hottest ones. For the coldest stars (K and M--type stars) with $T \le$ 3750 K we prefer to use models from Bessell {\it et al.} (1989, 1991a,b) for M Giants (for various Z) 
and from Brett (1995a,b) for M dwarfs, with a single metallicity 
Z=0.02. The initial helium content of the gas from which stars form is computed with an initial He fraction $Y = 0.24$ and assuming ${\Delta}Y/{\Delta}Z = 3$.

\subsection{Modelling the Cosmic Chemical Evolution }
\hspace{0.5cm}
	A model of chemical evolution allows the abundance evolution in the ISM and the abundance distribution in stars by solving a simple set of equations (Tinsley 1980). For a population of galaxies, the chemical evolution equations can be expressed in terms of comoving densities of stars, neutral gas and heavy--elements $\Omega_g$, $\Omega_*$, $\Omega_Z$ in unit of the critical mass density $\rho_c = 7 \times 10 ^{10}$ M$_{\odot}$ Mpc$^{-3}$ :

\begin{equation}
\frac{dZ\Omega_{g}}{dt}= - Z {\dot{\Omega}_{*}} + {\dot{\Omega}_Z} -{\dot{\Omega}^{Z}_{out}} 
\end{equation}
$Z$ is the mean abundance of the ISM of galaxies defined by the ratio $\Omega_Z$/$\Omega_g$. ${\dot{\Omega}^{Z}_{out}}$ is the comoving rate of outflowing metal--enriched gas out of galaxies. In the case of closed--box models this term vanishes.\\
 We use the $Z$--dependent yields of M92 with moderate mass loss for massive stars. For intermediate stars we use the Renzini and Voli (1981) yields. We do not make the assumption of instantaneous recycling but we track metals ejected by stars at each timestep and let new stars form out of the enriched gas, assuming instantaneous mixing of the metals.  
\subsection{Dust spectra}
\hspace{0.5cm}
  	A non--negligible part of the energy released by stars in the UV/optical wavelengths is absorbed 
by dust and re--emitted in the IR/submm. A normal spiral such as our Galaxy emits $\sim 30 \%$ of its light in the far--infrared (see Dwek {\it et al.} 1998 and reference therein). Taking dust obscuration into account is crucial for the census of stellar 
mass, metals and energy. Models of SEDs which correctly connect UV/vis and IR/submm windows are therefore needed. However modelling dust extinction is not an easy task since we have to 
address critical issues such as the dust properties, its geometrical distribution with respect to stars and its chemical evolution. Models of SEDs that take into account the extinction and emission effects of dust in a consistent way have been developed by several authors (Franceschini {\it et al.} 1991, 1994), Fall, Charlot and Pei (1996), Silva {\it et al.} (1998), Pei {\it et al.} (1999, hereafter PFH), but all these models assume a simple relation between dust and metal abundance of the gas and do not have a detailed modelling of the starlight re--processing by dust grains. More detailed and self--consistent models of SEDs called STARDUST have recently been developped by DGS. Here we will follow DGS to compute the global SED from UV/vis to IR/submm wavelengths.\\
\subsubsection{Dust absorption}
\hspace{0.5cm}
	The first step is to compute the optical depth. We will first assume a simple prescription given by: 

\begin{equation}
\tau_{\lambda}(t)=(\frac{A_{\lambda}}{A_{V}})_{{Z_\odot}} {\tau}_{V}
\end{equation}
\noindent
where $(A_{\lambda}/A_{V})_{\odot}$ is the {\it extinction curve} in our Galaxy (solar metallicity) from Mathis {\it et al.} (1983). In principle one should relate $\tau_{\lambda}(t)$ to the mean H--column density which is 
proportional to the mean gas density times the gaseous disk scale ($N_{H}(t)=M_{g}(t)/\pi r_{g}^{2}$ $\propto \rho_{gas} r_{g}$). To account for this, we introduce a ``fudge'' factor $\tau_{V}$.\\
As in Guiderdoni and Rocca--Volmerange (1987, hereafter GR\-V87), we assume a face--on ``slab'' geometry distribution where 
the gas and the stars which contribute mainly to dust heating are distributed with equal
height scales in the disks. Following GRV87, if $\tau_\lambda (t)$ is the optical thickness of
the disks at wavelength $\lambda$ and time $t$, the mean internal extinction 
correction (averaged over inclination angle $i$) is then given by:
\begin{eqnarray}
A_\lambda (t) &=& -2.5 \log < {1.-exp(-a_\lambda \tau_\lambda(t)/cosi)
\over a_\lambda \tau_\lambda(t)/cosi}>_i ,
\end{eqnarray}

The factor $a_\lambda \equiv (1-\omega_\lambda)^{1/2}$ takes into account the
effect of the albedo $\omega_{\lambda}$.  This ``slab'' geometry  is an intermediate case between the ``screen'' geometry, where the dust layer lies in front of the stars layer  and the ``sandwich'' geometry, where the dust layer is trapped between two star layers, geometries which leads respectively to a larger and a smaller absorption (Franceschini and Andreani 1995, Andreani and Franceschini 1996). \\
We have also to take into account the evolution of $\tau_V$ with redshift.  As already mentionned, $\tau_V$ is in principle related to the mean H--column density $N_{H}$ and depends on the geometrical structure of the gas. From scaling argument $\tau_{V} \propto \rho_{gas} r_g$ we would expect that $\tau_{V} \propto (1 + z)^{\alpha}$ with $\alpha = 2$. However, other factors might evolve, such as the properties of the dust or the actual geometry of the gas distribution within the disk so that the redshift--dependence might deviate from the simple scaling relation. In this case, the slope index $\alpha$ is taken as an adjustable parameter whose value is determined when consistency with the HDF rest--frame UV luminosities is obtained.\\
The above modelling of the mean optical thickness is very crude and probably does not reflect the complexity of dust extinction. In the following, we will use  the GRV87 prescription in which the mean optical thickness is explicitly related to the mean metallicity and to $N_{H}$. Following GRV87 and Franceschini {\it et al.} (1991) we will assume that the mean face--on optical depth of the average gas disk at time $t$ and wavelength $\lambda$ is given by :
\begin{equation}
\tau_\lambda(t)  = \left( {A_\lambda \over
A_\mathrm{V}} \right)_{Z_\odot} \left( {Z(t) \over Z_\odot} \right)^s 
\left({\langle N_{\sc\rm H}(t) \rangle \over
2.1~10^{21} {\rm~at~cm^{-2}}} \right) \, ,
\label{eqtau}
\end{equation} 
where the mean H column density (accounting for the presence of
helium) is written as:
\begin{eqnarray}
\langle N_\mathrm{H}(t) \rangle \simeq 
6.8~10^{21} g(t) f_\mathrm{H} \, \, \mathrm{atoms \, cm^{-2}} \, \, .
 \nonumber
\end{eqnarray} 
where g(t) is the gas fraction. A galaxy with $g \approx 0.2$ has $<N_{H}> \sim 1.4 10^{21}$ atom cm$^{-2}$, in good agreement with the observational value for late--type galaxies (GRV87). $f_\mathrm{H}$ is a parameter which allows us to increase the H column density and is close to unity for normal spirals. The extinction curve here depends on the gas metallicity $Z(t)$ according to power law interpolations based on the Solar Neighbourhood and the LMC and SMC with $s = 1.35$ for $\lambda < 2000$  $\rm\AA$  and $s = 1.6$ for $\lambda > 2000$  $\rm\AA$.  
The derived extinction curves depend strongly on the adopted geometry, here we restricted ourselves to the ``slab'' geometry; keeping in mind that it might introduce some uncertainties in the obscuration curves. It would be interesting to compare this with more detailed models of extinction curves (see for example Ferrara {\it et al.} 1999) but this is beyond the scope of this present work.\\

Finally, the observed flux $F^{*}_{o \lambda}$ is then related to the emitted (intrinsic) flux $F^{*}_{\lambda}$ by:\\

\begin{equation}
\frac{F^{*}_{o\lambda}}{F^{*}_{\lambda}}=10^{-0.4A_\lambda(t)}
\end{equation}

The bolometric luminosity released at IR/submm wavelengths is then given by:
\begin{equation}
L_{IR}(t) = \int F^{*}_{\lambda} (t) (1-10^{-0.4A_\lambda(t)})d\lambda .
\end{equation}
\subsubsection{Dust emission}
\hspace{0.5cm}
	In order to compute the IR/submm emission spectra, one needs to redistribute the total IR luminosity $L_{IR}$ over the whole range of wavelengths. This is achieved by using a dust model of the ISM. We assume the three--component (Polycyclic Aromatic Hydrocarbons, Very Small Grains and Big Grains) dust model 
developed by D\'esert {\it et al.} (1990) according to Maffei's method (Maffei 1994) which uses 
the observational correlations of the {\it IRAS} flux ratios with $L_{IR}$.\\

Three components are considered in D\'esert {\it et al.}  model spectra:\\

\begin{itemize}
\item{the Big grains (BG). The are made of silicates and graphite and have sizes between 10 nm and 0.1 $\mu$ m and are almost in thermal equilibrium. They are reasonably described by a modified blackbody $\epsilon _{\nu} B_{\nu}(T_{BG})$ with emissivity ${\epsilon_{\nu}}={\nu}^{m}$ (where the index $1 \le m \le 2$.}
\item{ Polycyclic aromatic hydrocarbons (PAHs). Because these molecules are small in size (less than 1 nm) they never reach thermal equilibrium when excited by the UV/vis radiation. Their temperature fluctuates and can exceed the equilibrium temperature which produces the 12 $\mu$m excess and the bands at 3.3, 6.2, 7.7 and 11.3 $\mu$m.}
\item{Very small grains (VSG) with sizes between 1 and 10 nm also made of silicates and graphite. As PAHs they never reach thermal equilibrium and therefore their emission spectrum is much broader than the black body spectrum at a given temperature.}

\end{itemize}

\begin{figure}[]
\centerline{\psfig{file=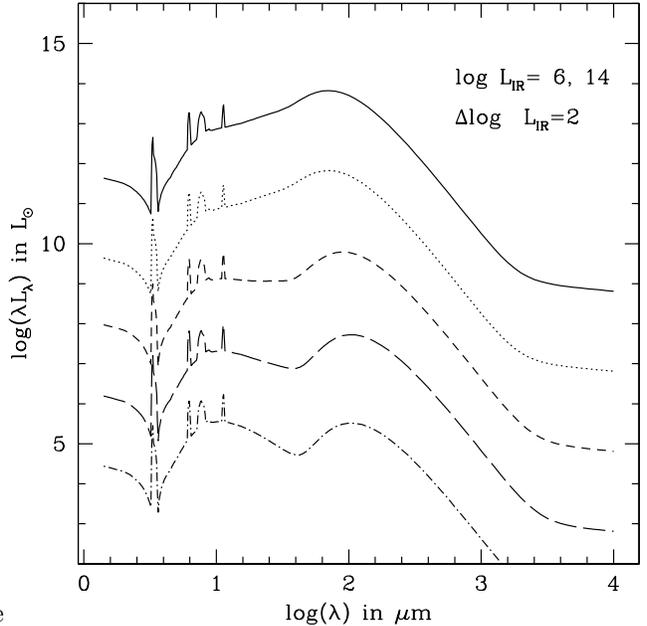,height=3.5in,width=3.5in}}
\caption{\small Model spectra in the IR/submm for different total IR luminosities ranging from $10^{6}$ to $10^{14}L_{\odot}$ and an emissivity index $m=1.5$.
\label{fig:LIR}}
\end{figure}

 The temperature T$_{BG}$ of the Big Grains is given by the IRAS 60/100 colours provided we fix the value of the emissivity index {\it m}. The derived temperature is between 15 and 50 K. Here we adopt the standard value of $m = 1.5$. Model spectra with $m =1.5$ seems to cover all the range of observations in the submm (Guiderdoni {\it et al.} 1998). The contributions of each component is then calculated iteratively from the 12/100, 25/100 and 60/100 ratios. The resulting spectra are computed from few $\mu$m to several mm and evolve with the total IR luminosity L$_{IR}$ such that galaxies with higher L$_{IR}$ emit preferentialy at shorter $\lambda$ (Fig. \ref{fig:LIR}). Here we will adopt the model spectra of dust emission with total IR luminosity of $L_{IR}=10^{10}L_{\odot}$ which is typical of normal galaxies and corresponds to a temperature of BG of  $T_{BG}\approx 16.5$ K. In spite of its limitations, this method is able to reproduce the FIR photometry of various individual galaxies (Maffei 1994).\\
\\
 
\begin{figure}[]
\centerline{\psfig{file=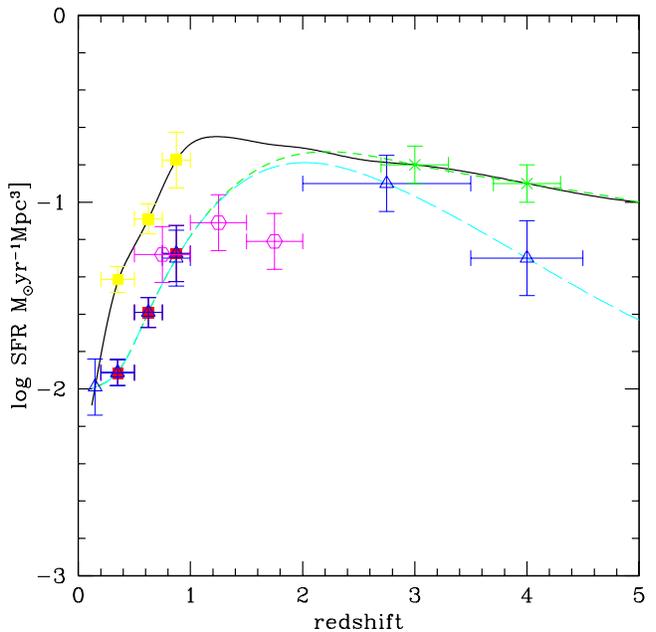,height=3.5in,width=3.5in}}
\caption{\small The star formation rate as function of redshifts. The curves are the different star formation histories we input in our models: ({\it solid line}), ({\it  small dashes}) and ({\it long dashes}) corresponding to $H$, $L_1$ and $L_2$ models. The data points are taken from M96 and F99 ({\it filled squares}): lower points and upper points respectively. Treyer {\it et al.} 1997 ({\it empty triangle}), Connolly {\it et al.} 1997 ({\it empty hexagons}). At high--$z$, the data are from Madau {\it et al.} 1998 ({\it open triangles}) and from S99 after correction for extinction ({\it crosses}).
\label{fig:sfr}}
\end{figure}


\subsection{Assumptions and inputs of the model}
\hspace{0.5cm}
	We choose three classes of CSFR. The high SFR model (model $H$) choosen in order to be consistent with the recent estimations based on IR surveys (F99) and large ground--based survey of Lyman--Break galaxies (S99). The low SFR models $L_{1}$ and $L_{2}$ are consistent with the UV/optical derived SFR with respectively a declining and a flat SFR at high redshifts (see table 1).  Fig. \ref{fig:sfr} shows the three models of CSFR we used, together with a compilation of the CSFR estimations derived from UV/vis and IR surveys.\\
In our modelling of dust obscuration we used the HDF rest--frame attenuated 1500  $\rm \AA$ emissivity at $z = 3-4$ (from S99 for models $H$ and $L_1$ and from M96 for model $L_2$) and the local emissivity at 100 $\mu$m to determine the two adjustable parameters, $\tau_V(z=0)$ and $\alpha$.\\
In the case of the closed--box approximation (i.e $\dot{\Omega}^{Z}_{out}=0$) the equations of chemical evolution (eq. 1) are solved assuming that at early times the stars form from some amount of gas whose initial gas density fraction is fixed by imposing that the present--day amount of HI gas is of the order of $\Omega_{g}(0) \approx 5 \times 10^{-4}$ (as estimated by Briggs \& Rao 1993). In the case where the closed--box approximation is relaxed, we will assume for simplicity a constant outflow ${\dot{\Omega}^{Z}_{out}}= const$. 
Note here that the ISM gas content of the galaxies is assumed to be mainly due to neutral hydrogen gas HI; we neglect both the ionized and molecular forms of the gas (HII and $H_2$). This approximation is reasonable at high redshifts, as measurements of gas content of Damped Lyman $\alpha$ systems show low ionization states (Viegas 1995) and low molecules abundances (Levshakov {\it et al.} 1992, Petitjean {\it et al.} 2000). Locally, this approximation may not hold since the present--day late--type galaxies contain on average a non--negligible fraction of $H_{2}$ (Young \& Scoville 1991) which would result in a factor of two uncertainty on $\Omega_{gas}$ or more if $H_2$ contribution to the ISM gas is more important (Pfenniger {\it et al.} 1994).\\
We adopt a Salpeter IMF $\phi(m) \propto m^{-x}$ with $x=1.35$ for $m_{d}\leq m \leq m_{up}$ and $m_{d}=0.1$ M$_{\odot}$ and $m_{up}=100$ M$_{\odot}$. This IMF has been found to be more reliable in reproducing the observed emissivities locally than the Scalo IMF (Lilly {\it et al.} 1996, M98).\\
Finally, we will suppose to a first order that the results are independent of the cosmological parameters. Discussion of the effect of the cosmological parameters on the luminosity density can be found in (Cass\'e {\it et al.} 1998).\\  
\begin{table}
\begin{center}
\begin{tabular}{clclcl|}
\hline
model    &       $0 \le z \le 1$        &  $z \ge 2$   \\ \hline\\

$H$	 &  consistent with F99 & consistent with S99\\

$L_{1}$  & consistent with M96 & consistent with S99\\

$L_{2}$ & consistent with M96  & consistent with M96\\
\hline
\end{tabular}
\caption{\small Definition of the class of models describing the CSFR histories we input in our modelling. $H$ model is chosen to be consistent with the mid--infrared CSFR inferred from observations of ISO 15 $\mu$m selected galaxies in the CFRS at $0 \le z \le 1$ by F99 and with the CSFR inferred from large ground--based survey of Lyman--break galaxies by S99 and corrected for extinction at $z > 2$ . Models $L_{1}$ and $L_{2}$ both are consistent with the CSFR derived from the CFRS survey by M96 at $0 \le z \le 1$, at high--$z$ model $L_{1}$ is consistent with S99 while $L_{2}$ is consistent with M98
\label{table-1}} 
\end{center}
\end{table}

\begin{figure}[hbtp]

\psfig{file=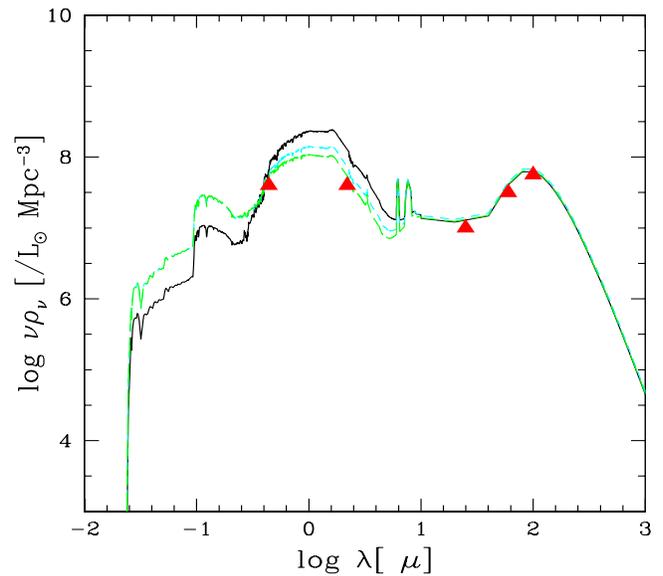,height=3.1in,width=3.5in}

\caption{\small The rest--frame spectral energy distribution SED from UV/vis to FIR/submm at $z = 0$. The data points are from Ellis {\it et al.} (1996) for 0.44 $\mu m$ ; Gardner {\it et al.} (1997) for 2.2 $\mu m$; and Soifer \& Neugbauer (1991) for 25, 60 and 100 $\mu m$.
\label{fig:rho_univ1}}

\end{figure}

\section{The results and comparison to observations}
\hspace{0.5cm}
Deep galaxy surveys have allowed the probing of rest--frame UV/optical to NIR emissivities of high and low redshift galaxies. When combined with the CIB measurement it provides us with crucial constraints on the global history of star formation.\\

\subsection{The rest--frame comoving emissivity from UV/optical to IR/submm}
\hspace{0.5cm}
	We are interested in the global evolution of the radiation from both the stellar and the dust component. This can be described by the comoving luminosity density $\rho_{\nu}$ as a function of redshift defined as the energy radiated  per unit frequency per unit of co--moving volume at redshift $z$. The intrinsic stellar emissivity can be expressed in terms of the global star formation rate $\dot\Omega_{*}$
\begin{equation}
\rho^{*}_{\lambda}=\rho_{c} \int^{t}_0 F^{*}_\lambda(t-t') \dot\Omega_{*}(t'){\rm d}t'
\end{equation}
where $F^{*}_{\lambda}(t-t')$ is the stellar population spectrum as computed by our population synthesis code (see DGS for details).\\
The spectral luminosity density at each redshift is computed by simply summing the observed (attenuated) stellar radiation and the dust radiation: $\rho_{\nu}(z)=\rho^{*}_{o\lambda}(z) + \rho^{d}_{\lambda}(z)$ where $\rho^{*}_{o\lambda}$ is related to the intrinsic (unabsorbed) emissivity by:
$\rho^{*}_{o\lambda}= 10^{-0.4A_\lambda} \rho^{*}_{\lambda}$.\\
We first fix the adjustable parameters of the models $\alpha$ and $\tau_{V}$, by fits to the observed luminosity density at 1500 $\rm \AA$ at $z=3-4$ and the local to the local 100 $\mu$m emissivity respectively. The best--fit solutions are obtained for the following values of the parameters ($\tau_{V}$, $\alpha$): (1.3, 1.4), (1.2, 1.4) and (1.2, 0.8). The corresponding values of the extinction factor $A_{V}$ varies from $A_{V}=0.45$ locally to $A_{V}=2-2.5$ at $z=3-4$ for model $H$. In the case of model $L_2$,  $A_{V}$ is almost constant, varying from 0.45 locally to 0.53 at $z=3-4$. Fig. \ref{fig:rho_univ1} shows the resulting $z = 0$ co--moving spectral luminosity density from UV/vis to IR/submm wavelengths, on which are superimposed the data. Our models with a Salpeter IMF give overall good fit to the present--day stellar light from UV/vis to IR. Except for the 2.2 $\mu$m emissivity which seems to be overestimated by $H$--type models, our models are able reproduce the data not only at 100 $\mu$m (which is expected since it is used as an input) but also at $\lambda=0.44$, 2.2, 25, 60 $\mu$m, whatever the chosen CSFR history. This result shows that local co--moving emissivities are not sensitive to the CSFR and therefore are not suitable to constrain the star formation rate history. We also compute the resulting spectral luminosity density in the case where GRV87 prescription for $\tau_{\lambda}(t)$ is used and found no difference with models which use our simple prescription (eq. 2); the two SEDs are almost undistinguishable.\\
We use for the determination of dust emission a model with $L_{IR}=10^{10}L_{\odot}$. To take into account the uncertainties in the FIR part of the SED, we run our models with a higher IR luminosity $L_{IR} \ge 10^{12}L_{\odot}$ which is typical of the Ultra--Luminous Infrared Galaxies (ULIRGs). As expected, we found that the peak emission in Far--Infrared (100 $\mu$m) is shifted toward shorter wavelengths, leading to a poorer fit to the data. \\

\subsection {The Extragalactic Background Radiation}
\begin{figure}[hbtp]
\begin{minipage}{65ex}
\psfig{file=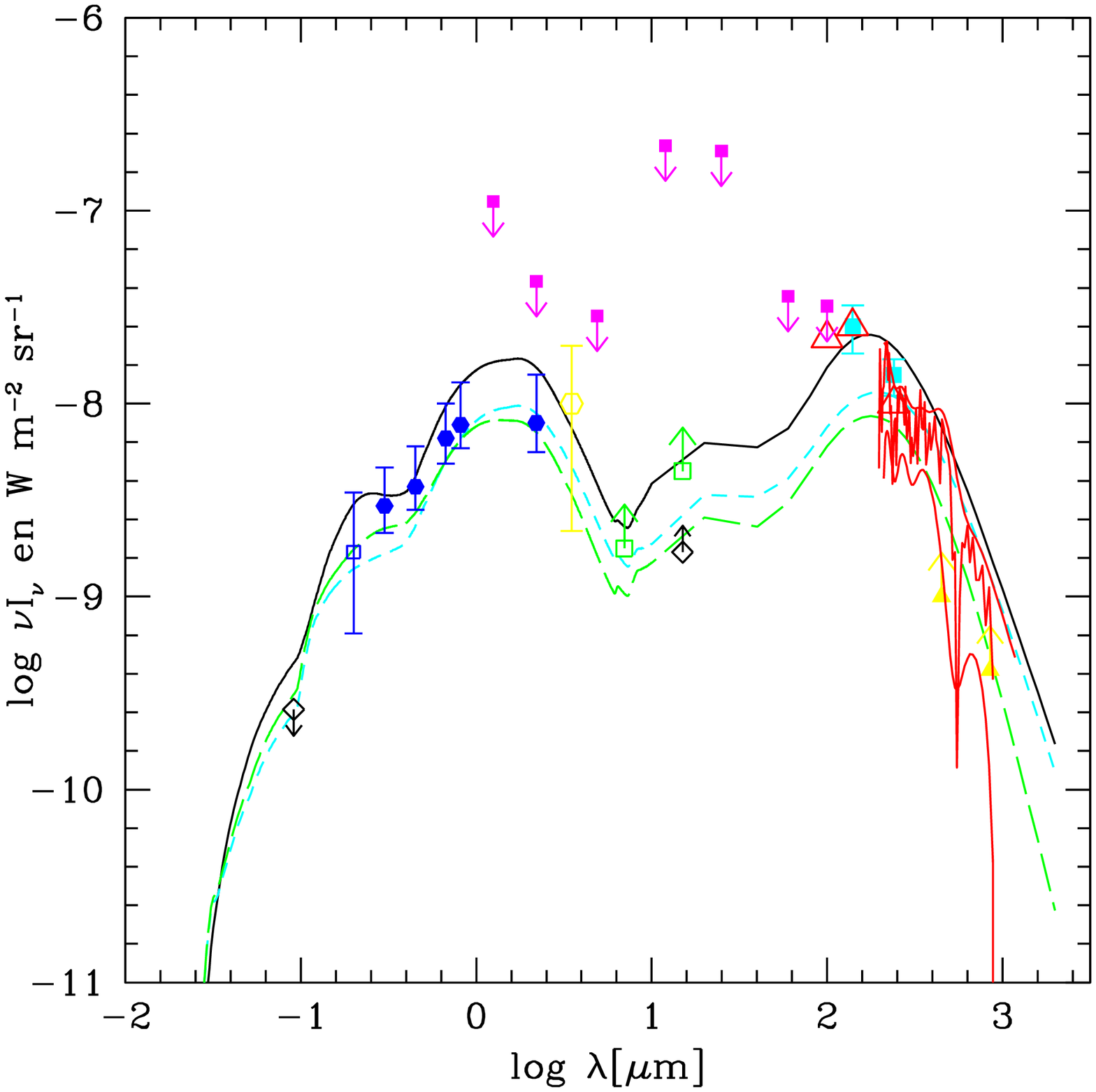,height=3.1in,width=3.5in}
\end{minipage}
\begin{minipage}{65ex}
\psfig{file=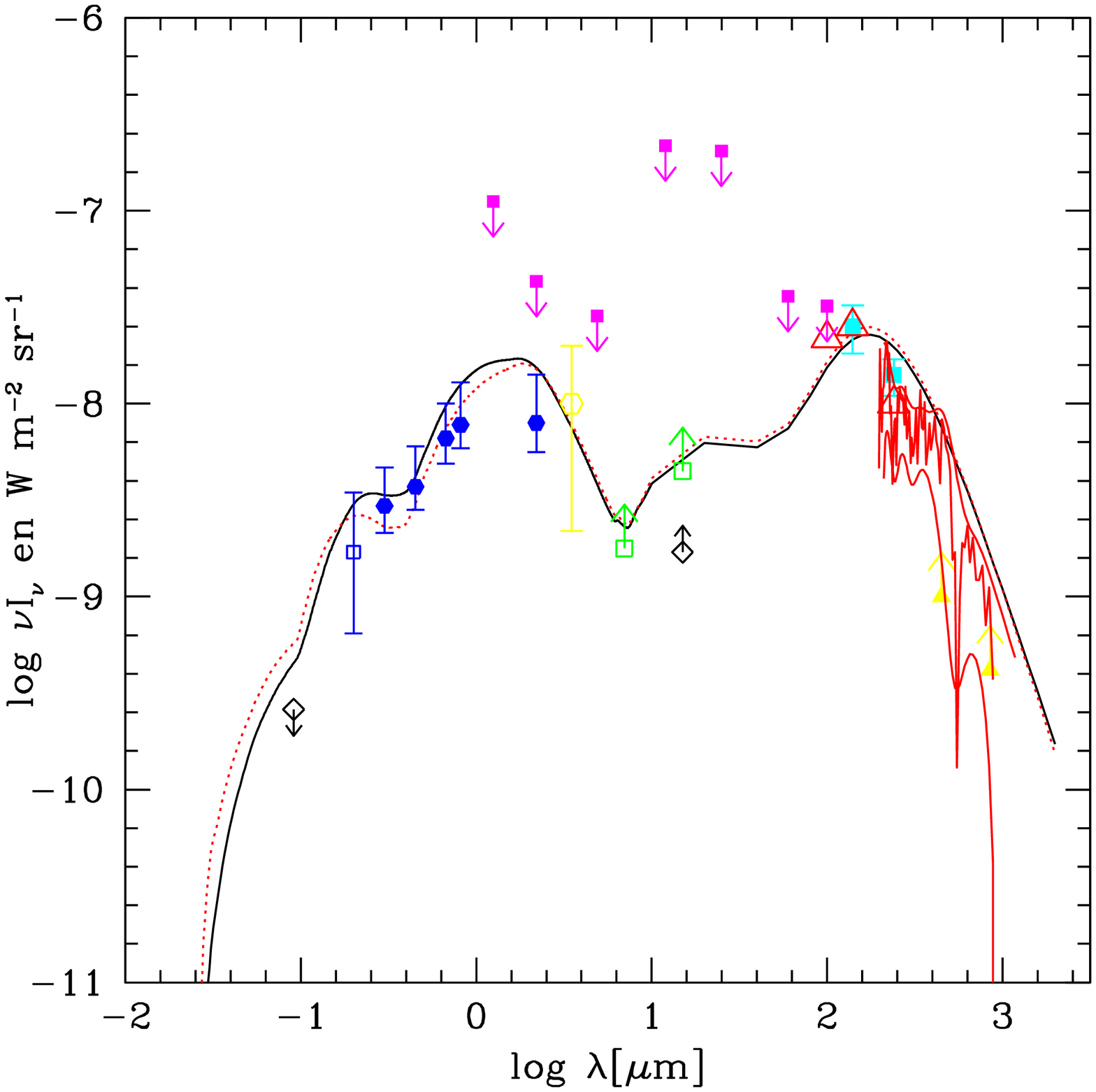,height=3.1in,width=3.5in}
\end{minipage}
\caption{\small The predicted diffuse backgrounds in the optical and FIR/submm as compared to current limits and detections. The coding of the curves are same as in fig 1. {\it Solid hexagons} result from the faint galaxy counts at 3600 - 22000  $\rm\AA$ (Pozzetti {\it et al.} 1998). The {\it empty square} results from galaxy counts at 2000  $\rm\AA$  (Armand {\it et al.} 1994). The {\it Empty hexagon} is the detection at 3.5 $\mu$m from COBE/DIRBE by Dwek \& Arendt (1998). The {\it filled squares} correspond to the COBE/DIRBE residuals (Hauser 1996) only 140 $\mu$m and 240 $\mu$m signals are firm detections (Hauser {\it et al.} 1998), all other points are upper limits. {\it Empty triangles} are the 100, 140 and 240 $\mu$m estimations from Lagache {\it et al.} (2000). The {\it Filled triangles} are the estimations of the CIB at 850 and 450 $\mu$m from repsectively Blain {\it et al.} 1999) and Smail {\it et al.} (1997) updated by Ivison {\it et al.} (1999). The {\it Empty squares} are the lower limit to the background at 7 and 15 $\mu$ m from Altieri {\it et al.} ( 1999). The {\it empty losanges} are the lower limit to the 15 $\mu$m and an upper limit to the 0.912 $\mu$m background by Vogel {\it et al.} (1995). The zigzag is the COBE/FIRAS residuals detected by Puget {\it et al.} (1996) and revisited by Guiderdoni {\it et al.} (1997). The lower panel shows the predicted diffuse background intensity obtained using GRV87 prescription of dust extinction ({\it dotted line}) as compared to the one derived using our simple prescription ({\it solid line}).  \label{fig:fond1} }
\end{figure}

\hspace{0.5cm}
	The second observational constraint is set by the extragalactic background light (EBL). Cosmological implications of the CIB on the CSFR have been discussed by several groups, for example M98, Dwek {\it et al.} (1998), Gispert {\it et al.} (2000). Here we use our models to derive the intensity of the background. The intensity $I_{\nu}$ of EBL results from integrating the emission coming from various extragalactic objects on the line of sight over the cosmic time. \\
The intensity $I_{\nu}$ at a given time t is given by the following:\\
\begin{equation}
I_\nu={c\over4\pi}\int^{\infty}_0 dz \rho_{\nu(1+z)}\biggl|{dt\over
dz}\biggr|.
\end{equation}
We here ignore the contribution of AGNs in heating the dust. 
Predictions of our models for the intensity of the background radiation in the optical and IR/submm \footnote{Recent tentative of CIB detection at 2.2 and 3.5 $\mu$m has been reported by Gorjian {\it et al.} (2000), our best--fit model is consistent with their estimated intensity at 2.2 and 3.5 $\mu$m} are shown in Fig. \ref{fig:fond1}. The overall spectrum seems to be well reproduced by all the models regardless of their very different SFR histories. In particular, the shape of the submm part of the extragalactic background does not discriminate between the models, and is therefore not a very crucial probe of the star formation history. Instead, the CIB data point at 140 $\mu$m is very discriminant. Only mid--infrared derived CSFR at low--$z$ (model $H$) is able to correctly match this point while UV--derived low--$z$ CSFR ($ L_{1}$ and $L_{2}$ models) , falls considerably short of the COBE detections at $\lambda=140$ $\mu$m. Moreover, the CIB is less sensitive to the CSFR shape at high--$z$. For example, had we chosen a model similar to $H$--model at low--$z$ but with a declining SFR at high--$z$, we would obtain an accpetable fit to the 140 $\mu$m data. This result suggests that the bulk of the energy in the FIR/submm is due to moderately obscured normal galaxies lying at low and 
moderate redshifts ($0\leq z \leq 1$) and not by a population of heavily extinguished star--forming regions located at very high redshifts, as previously claimed (Blain {\it et al.} 1998, hereafter B98). We also show the predicted intensity of the background radiation when the GRV87 prescription for the optical depth $\tau_{\lambda}$ is used (eq. 4). As we can see there is no strong difference between the two prescriptions.\\
It is interesting to compare our result with other works. PFH have made use of many observational inputs (from QSOs absorption line surveys, optical imaging, redshift surveys, and FIRAS/DIRBE cosmic infrared background) to derive solutions for the cosmic histories of star formation rate. They found a global star formation rate which is higher than the UV--derived one, in good agreement with our result. Although their three solutions do not fit the CIB at 140 $\mu$m, their solutions are actually significantly below this data point (see their fig. 11). Our result in the other hand, is in disagreement with previous works in which it is claimed that the best CSFR solution is higher and flatter at high redshifts than the optically derived form (B98, Gispert {\it et al.} 2000). Note that these authors used a different approach than ours and did not use the point at 140 $\mu$m to discriminate between the models, as in the present work.\\

\section{Cosmological history of metals}
\hspace{0.5cm}
	In order to get a consistent picture of the evolution of the universe, one should relate the CSFR to the metal enrichment of the universe and its stellar content. To achieve this goal we will use the CSFR model that account for all available data from UV/vis to FIR/submm wavelengths to compute the enrichment history of the universe. 
	
\subsection{Chemical enrichment of the ISM}
\begin{figure}[!h]
\psfig{file= 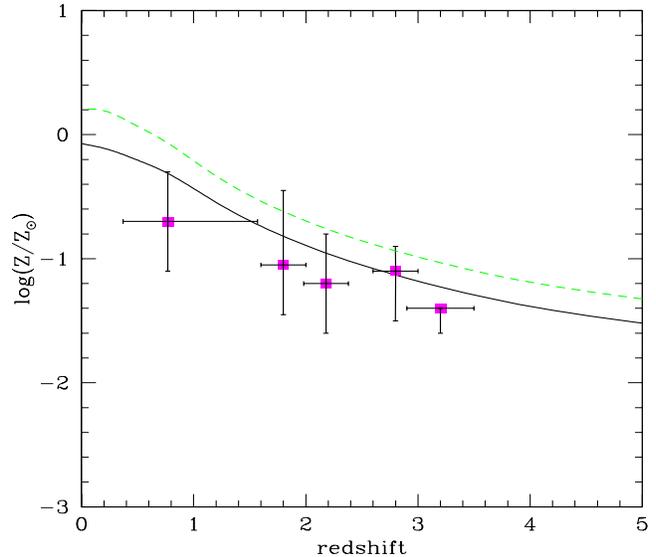,height=3.1in,width=3.5in}
\caption{\small Evolution with redshift of the mean comoving ISM metallicity as predicted by the best--fit model ({\it solid line}). The {\it dotted line} curve corresponds to the case of closed box evolution. The data points correspond to the metallicity as inferred from the DLAs Pettini {\it et al.} (1997). The point at $z = 0.77$ is from Boiss\'e {\it et al.} (1998).
\label{fig:ZISM}}
\end{figure}

\hspace{0.5cm}
	Our knowledge of chemical evolution was based on local observations 
(mainly the solar neighbourhood) until the outcome of high--$z$ observations 
of damped Ly$\alpha$ systems, which allowed us to draw some conclusion about galaxy evolution in terms of the cosmic evolution of the mean metallicity and cool neutral gas. There are some indications that these DLAs represent the progenitors of present--day galaxies, although the nature of these objects is still a matter of debate. Moreover, it is not yet clear whether the mean HI weighted metallicity derived from current observations of DLAs is representative or not of the global metal enrichment of the Universe. The picture that emerges from the DLAs observations is that {\it i}) the neutral gas content is decreasing from $z \sim 3$ to the present--epoch, corresponding to the conversion of gas into stars and that {\it ii}) the metallicity is decreasing with increasing redshift, with very low metallicities at high redshifts, which would suggest that these systems are in their early phase of enrichment. \\
Figure \ref{fig:ZISM} shows the mean abundance of metals in the interstellar medium $Z_{ISM}$ as function of redshifts compared to the metallicity at high redshift derived from the observed abundance of Zn in damped Ly$\alpha$ systems (Pettini {\it et al.} 1997), which should agree if DLA and field galaxies have the same metal enrichment. The uncertainty in the data is still large though, and no attempt has been made to correct for the observational biases which could potentially alter our interpretation of the DLAs metal abundances (Prantzos \& Boissier 2000 and reference therein) but see Savaglio 2000. We find that in the closed--box assumption, our best--fit to the CIB at 140 $\mu$m model clearly fails to reproduce the data; it overpredicts the mean metallicity by a factor of $\sim 3$ at $2 \le z \le 3$. This is less than that found by B98 with their submm--derived CSFR (a factor as high as 5). Our best--fit model also produces a mean metallicity $Z(0)$ which is $\approx 2$ times larger than the metallicity in the local universe if we assume it to be solar. \\ We have therefore relaxed the closed--box approximation and allowed for some outflowing of metal--enriched gas from the galaxies. As we can see from figure \ref{fig:ZISM}, the agreement with the data becomes better than in the closed--box models. If we further assume that the ejected metals are responsible for IGM pollution, we can use our model to estimate the amount of metals present in the IGM.\\
\subsection{Chemical enrichment of the IGM}
\hspace{0.5cm}
The detections of CIV and Si IV absorption lines in the Ly$\alpha$ forest as reported (Tytler {\it et al.} 1995, Songaila \& Cowie 1996, Songaila (1997) can potentially give us information about the early enrichment of the intergalactic me\-dium (IGM) with metals (Gnedin \& Ostriker 1997). If we assume that the IGM metal content can well be represented by the metallicity inferred from Ly$\alpha$ clouds, we may compare our prediction of the density of metals to the observations of CIV and Si IV absorption lines in the Ly$\alpha$ forest. Assuming that ejected metal--enriched gas is uniformely distributed in the inter\-galactic medium, we estimate the mean co--moving density of metals in the IGM by integrating $\dot{\Omega}^{Z}_{out}$ over the time. We find that the density of heavy--elements at $z = 0$ is $\Omega^{IGM}_m=7.7 \times 10^{-5}$ (i.e $\rho_{m}=5.4 \times 10^{6}$ M$_{\odot}$ Mpc$^{-3}$). At $z \approx 3.2$, $\Omega^{IGM}_m \sim 6 \times 10^{-7}$ in agreement within the error bar with the data point of $\Omega^{IGM}_m \sim 8 \times 10^{-7}$ reported by Songaila (1997). This result supports the general belief that the chemical enrichment of the IGM may be due to the outflowing of metals from galaxies, although this scenario is still a matter of debate (see Gnedin 1997 for an alternative scenario). \\
It has been pointed out that if no differences exist between field and clusters galaxies, then clusters of galaxies can provide us with information about the chemical enrichment history of the universe (Renzini 1997). We therefore estimate the local IGM mean metallicity $Z_{IGM}(0)=\Omega^{IGM}_m/\Omega_b$ and find it to be of the order of $10^{-3}$. Using the low value of D/H ( $\eta_{10} \sim 5.3$) by Burles \& Tytler (1998) ($\Omega_{b}\approx 0.077$) and assuming a solar metallicty of 0.016, this value yields to a global mean metallicity of 6\% solar which is $\sim 4$ times smaller than the metallicity derived from X--ray observations of the intra--cluster medium ICM ($\sim 0.25Z_{\odot}$). As suggested by Renzini (1997), this cluster--field difference may be due to either cluster--related processes such as stripping or ram pressure or to a flatter IMF in clusters relative to field galaxies. Note that if we use the high value of D/H $\approx 10^{-4}$ (e.g. Lemoine {\it et al.} 1999 for a review) yielding $\Omega_{b}\approx 0.02$, we would obtain a global metallicity of 0.2 solar which is close to the ICM metallicity.\\ Regarding the large uncertainties in the observations and the limitations of our models (a constant metal--enriched gas ejection is certainly a crude approximation), we should be cautious before drawing any conclusion regarding the global enrichment of the universe.\\

\subsection{Stellar content of the universe}
\begin{figure}[h]
\centerline{\psfig{file= 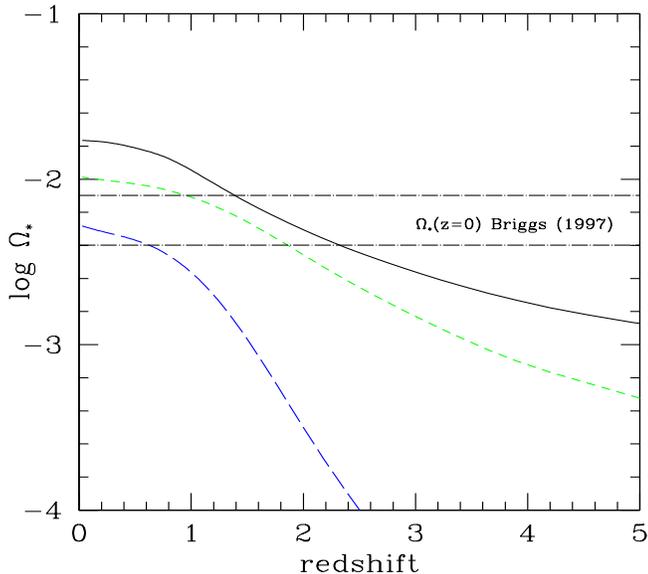,height=3.1in,width=3.5in}}
\caption{\small  Evolution as a function of redshift of of the co--moving stellar mass density. Coding of the models are as in fig. \ref{fig:sfr}. The horizontal lines define the 
data of the total mass density of the stars at the present epoch as estimated
 by Briggs (1997)
\label{fig:star}}
\end{figure}

\hspace{0.5cm}
	Our models can also provide the global evolution of the stellar content. Fig. \ref{fig:star} shows the inferred evolution of the stellar content as a function of the cosmic epoch for each of the models of CSFR . The estimate of the current $z = 0$ $\Omega_{*}(0) \sim 0.004-0.008$ comes from Briggs (1997). We can see that the best--fit model to the 140 $\mu$m data point predicts a present--day stellar mass density larger by at least a factor of two than the current value. This is very important as it shows that models able to produce enough energy in the FIR part of the CIB lead to an excess of stars as compared to the present--day estimates. Note that the FIR/submm derived CSFR of B98 produce values of $\Omega_{*}(0)$ at least a factor of two higher than that which our models produce; this is due to their higher CSFR at high redshifts. The current $\Omega_{*}(0)$ estimate is still uncertain though. There are several sources of errors that can bias the inferred $\Omega_{*}(0)$, such as the uncertainties in the derived luminosity functions parameters $\Phi^{*}$ and $L^{*}$; on the estimation of the contribution of starlight from low surface brightness (LSB) galaxies; in the adopted stellar M/L ratio of disks, spheroids and irregulars. However all these effects are not very important and cannot account for the discrepancy. More exotic solutions should therefore be invoked. Several solutions are possible, first it is possible that some stars (extragalactic stars, machos) are missed in the census of the stellar budget but it is unlikely that we missed that significantly.\\
Second, contamination by AGN causes problems that are more important than we think. We know from IRAS and SCUBA observations and recently from X--ray {\it Chandra} observations (Barger {\it et al.} 2000) that a population of dust enshrouded AGNs do exist (Sanders and Mirabel 1996). However it is still unclear how important this contamination is as it is difficult to discriminate between a pure starburst and a dusty AGN in the object identification process. Therefore, it is highly probable that a large fraction of the CIB may be due to the dust enshrouded AGN population. AGN contamination also complicates the derivation of the star formation rate history from the observed luminosities even in the multi--wavelengths surveys (as the one derived by F99), although it is less critical than in the UV wavelengths.\\
Third, our adopted Salpeter IMF is responsible for such a discrepancy. A IMF biased against low--mass stars, would reduce the number of long--lived stars and thus attenuate the discrepancy. Evidence of a shallow mass function below 1 M$_{\odot}$ has been observed in the Galactic disk (Gould {\it et al.} 1996) and more recently in the Galactic bulge (Zoccali {\it et al.} 2000). A heavily biased IMF has also been suggested from microlensing data (Chabrier 1999).\\

\section{Conclusions}
\hspace{0.5cm}
	We have used a series of self--consistent chemo--spectro\-photometric models which take into account the metallicity and dust obscuration effects to study the glo\-bal evolution of spectrophotometric properties together with the stellar, chemical and radiation content of the galaxies averaged over a large co--moving volume, independently of their individual details and properties. \\

$\bullet$ We find that our models with a Salpeter IMF reproduce fairly well the observed present--day comoving luminosity density at $\lambda = 0.44\mu$m and 2.2 $\mu$m although mid--infrared derived CSFR ($H$--model) appears to overproduce the light at $\lambda = 2.2 \mu$m. At wavelengths dominated by the dust emission (at $\lambda= 25$, 60 and 100 $\mu$m), we find good agreement with the observations independent of the choosen CSFR. \\  

$\bullet$We show that the point at 140 $\mu$m is crucial for constraining the CSFR history and seems to favour a higher comoving star formation rate at low redshifts than the UV/optically derived one; however the shape at high redshifts of the CSFR is not well constrained by the CIB measurements, contrary to previous claims.\\

$\bullet$ We suggest that the main contribution to the bulk of the CIB and more specifically to the energy at 140 $\mu$m is due to normal galaxies, lying at low and moderate redshifts, and not to distant and dusty galaxies as previously suggested.\\

$\bullet$ Concerning the history of the metal enrichment, we find that in the frame of closed--box evolution, our best--fit model to the CIB at 140 $\mu$m overpredicts the metallicity as observed in DLAs. A better agreement is obtained in the case of a model with some outflowing metal--enriched gas.\\

$\bullet$ Assuming that the ejected metal--enriched material is responsible for the IGM metallicity we derive, the IGM metal content and found it to be consistent within the error bar with the metals in the Ly$\alpha$ forest. However, we cannot derived conclusions regarding the IGM metal--enrichment given the large uncertainties in the data and the crudness of our metal enrichment treatment. \\

$\bullet$ Our best--fit model to the CIB at $\lambda =140$ overpoduces the present--day stellar mass density $\Omega_{*}(0)$ by a factor of at least $\sim 2$ as compared to the current value. If the CIB at $\lambda =140$ $\mu$m is correct, then this discrepancy and the excess of light at $\lambda =2.2$ $\mu$m could be explained by our choice of Salpeter IMF. This would then be an indication in favour of an IMF which is biased toward massive stars, opening a new debate on the universality of the initial mass function. However, one may keep in mind that our calculations have been done under the assumption that the FIR/submm background is due to a stellar component only; we neglect any contribution of AGNs in dust heating, which is probably not true.     \\

\begin{acknowledgements}
We are very gratefull to A. Blanchard for fruitful discussions. We thank the referee A. Ferrara for his useful comments which helped us to improve the content of this paper. RS acknowledges support from AAS Chretien Grant.
\end{acknowledgements}


\begin{thebibliography}{}

\bibitem[]{Armand}
Altieri {\it et al.} 1999, A\&A 343, L65

\bibitem[]{Andreani}
Andreani, P. and Franceschini, A., 1996, MNRAS, 283, 85

\bibitem[]{Armand}Armand, C., Milliard, B., \& Deharveng, J.M., 1994, A\&A 284, 12

\bibitem[]{}
Aussel, H.; Cesarsky, C. J.; Elbaz, D. and Starck, J. L. 1999, A \& A 342, 313

\bibitem[1998]{Barger}
Barger A.\,J., Cowie L.\,L., Sanders D.\,B., Fulton E., Taniguchi Y., 
Sato Y., Kawara K., Okuda H., 1998, Nature  394, 248

\bibitem[2000]{Barger}
Barger, A.J., Cowie, L.L., Mushotzky, R.F., Richards, A.E., astro-ph/0007175

\bibitem[1989]{Bessel1}
Bessell, M.S., Brett, J.M., Scholz, M., \& Wood, P.R., 1989 A\&AS 77, 1

\bibitem[1991]{Bessel2}
Bessell,M.S, Brett, J.M, Scholz, M. and Wood, P.R, 1991a, A\&AS 89, 335

\bibitem[1991]{Bessel2}
Bessell,M.S, Brett, J.M, Scholz, M. and Wood, P.R, 1991b, A\&A 244, 251

\bibitem[1998]{Blain}
Blain A.W., Smail I., Ivison R.J. and Kneib J.-P., 1998, MNRAS 302, 632

\bibitem[1999]{Blain}
Blain A.W., Smail I., Ivison R.J. and Kneib J.-P., 1999, ApJ 512, L87 

\bibitem[]{}
Boiss\'e, P., Le Brun, V., Bergeron, J., \& Deharveng, J.M., A\&A 333, 841

\bibitem[1998]{Bretta}
Brett,J.M., 1995a, A\&A 295, 736

\bibitem[1998]{Brettb}
Brett,J.M., 1995b, A\&AS 109, 263

\bibitem[]{Briggs} 
Briggs, F. 1997, Publ. Astron. Soc. Australia 14, 31 

\bibitem[1998]{}
Burles, S. \& Tytler, D. 1998, ApJ 499, 699

\bibitem[]{}
Cass\'e, M., Olive, K.A., Vangioni-Flam, E. and Audouze, J., 1998, New Astr., 3, 259

\bibitem[1999]{Chabrier}
Chabrier, G., 1999, ApJL, 513, 103


\bibitem[1996]{Charbonnel}
Charbonnel, C., Meynet, G., Maeder, A.,Schaerer, D.1996, A\&AS 115, 339

\bibitem[1997]{Connolly}
Connolly, A.J., Szalay, A.S., Dickinson, M.E., SubbaRao, M.U., 
\& Brunner, R.J. 1997 ApJ 486, L11

\bibitem[]{}
 D\'esert, F.X., Boulanger, F., Puget, J.L. 1990, A\&A 237, 215

\bibitem[]{}

Devriendt, J.E.G., Guiderdoni, B. and Sadat, R. 2000, A\&A 350, 381 

\bibitem[]{}
Dwek, E., Arendt, R.G., Hauser, M.G., Fixen, D. {\it et al.}, 1998, ApJ 508, 106.

\bibitem[]{}
Dwek, E. and Arendt, R.G., 1998, ApJ 508, L9


\bibitem[]{}
Eales, S., Lilly, S., Gear, W. {\it et al.}, 1999, ApJ 515, 518

\bibitem[]{}
Elbaz, D. et al. 1999, A \& A 351, L37-L40

\bibitem[]{}
Ellis, R.S., Colless, M, Broadhurst, T., Heyl, J., Glazebrook, K., 1996, MNRAS 280, 235
\bibitem[]{}
Fall, S.M., Charlot, S., and Pei, Y.C., 1996, ApJ 464, L43

\bibitem[]{}
Ferrara, A., Bianchi, S., Cimatti, A. and Giovanardi, C., 1999, ApJS 123, 437
\bibitem[1999]{F99}
Flores, H., Hammer, F., {\it et al.}, 1999, ApJ 517, 148

\bibitem[]{Franceschini}
Franceschini, A. and Andreani, P., 1995, ApJ, 440, L5

\bibitem[1991]{Franceschini}
Franceschini, A., De Zotti, G., Toffolatti, L., Mazzei, P., and
  Danese, L., 1991, A\&AS 89, 285

\bibitem[1994]{Franceschini}
Franceschini, A., Mazzei, P., De Zotti, G., and Danese, L., 1994, ApJ 427, 140

\bibitem[1997]{}
Gardner, J. P., Sharples, R. M., Frenk, C. S., Carrasco, B. E. 1997, ApJ, 480, 99
 
\bibitem[2000]{}
Gispert,R., Lagache, G., \& Puget, J.L., 2000, A\&A 300, 1. 

\bibitem[1997]{Gnedin}
Gnedin, N.Y. and Ostriker, J., 1997, ApJ 486, 581

\bibitem[1997]{Gnedin}
Gnedin, N.Y., 1998, MNRAS 294, 407
\bibitem[1996]{Gould}
Gould, A., Bahcall, John N., Flynn, C., 1996, ApJ 465, 759G
\bibitem[2000]{}
Gorjian, V., Wright, E.L., \& Chary, R., R., 2000, ApJ 536, 550 

\bibitem[1987]{GRV87}
Guiderdoni, B \& Rocca-Volmerange, B., 1987, A\&A 172, 27

\bibitem[1997]{Guiderdoni}
Guiderdoni, B., Bouchet, F., Puget, J.L., Lagache, G. and Hivon, E. 1997, Nature 390, 257

\bibitem[1998]{Guiderdoni}
Guiderdoni, B., Hivon, E., Bouchet, F.~R., and Maffei, B., 1998, MNRAS 295, 877

\bibitem[1996]{H96}
Hauser M.\,G., 1996, in Kafatos M., Kondo Y. eds, Examining the Big Bang and Diffuse Background Radiations. Proc. IAU 168, Kluwer, Dordrecht 

\bibitem[1998]{H98}
Hauser M.\,G. {\it et al.}, 1998, ApJ 508, 25

\bibitem[]{} 
Hughes, D.H., Serjeant, S., Dunlop, J.S., Rowan-Robinson, M, Blain, A {\it et al.} 1998, Nature 
394, 241

\bibitem[]{}
Ivison, R., Smail. I., Blain, A., Kneib, J.P., astro-ph/9901361

\bibitem[1997]{Kawara}
Kawara K. {\it et al.}, 1997, in Wilson A.\ ed.,\ The Far-Infared and Submillimetre 
Universe. ESA publications, Noordwijk, p.\,285

\bibitem[1992]{Kurucz}
Kurucz, R.L., 1992 in the Stellar Populations of Galaxies, IAU Symp. 149,
ed. B. Barbuy A\&A Renzini, 225

\bibitem[]{} Lagache, G., Abergel, A., Boulanger, D\'esert, F.X., and Puget, J.L., 1999, astro-ph/9901059

\bibitem[]{}Lemoine, M. {\it et al.}, 1999, New Astr. 4, 231
 
\bibitem[]{} Lanzetta, K.M., Wolfe, A.M., Turnshek, D.A., 1995, ApJ 440, 435

\bibitem[]{}Levshakov, S. A., Chaffee, F. H., Foltz, C. B., Black, J. H. 1992, A\&A 262, 385

\bibitem[1995]{Lilly}
Lilly, S.J., Le F{\'e}vre, Hammer, F., Crampton, D., 1996, ApJ 460, L1
ApJ 455, 108

\bibitem[1996]{Madau1}
Madau, P., Ferguson, H.C., Dickinson, M.E., Giavalisco, M., 
Steidel, C.C., \& Fruchter, A. 1996 MNRAS 283, 1388 (M96)

bibitem[1998]{Madau1}
Madau, P., Pozzeti, L., Dickinson, M., 1998 ApJ. 498, 106 


\bibitem[1992]{Maeder} 
Maeder, A., 1992, A\&A 264, 105 

\bibitem[1994]{Maffei}
Maffei, B., 1994, {\it Ph.D. thesis}, Universit\'e de Paris VII

\bibitem[1983]{Mathis}
Mathis, J.S., Mezger, P.G, \& Panagia, N., 1983, A\&A 128, 212

\bibitem[1997]{Meurer}
 Meurer, G.~R., Heckman, T.~M., Lehnert, M.~D., Leitherer, C., \&
Lowenthal, J. 1997, AJ 114, 54 


\bibitem[]{PF}
Pei Y.\,C., Fall S.\,M., 1995, ApJ 454, 69

\bibitem[]{PF}
Pei Y.\,C., Fall S.\,M. and Hauser M. \,G. 1999, ApJ 522, 604

\bibitem[]{Petitjean}
Petitjean, P., Srianand, R., and Ledoux, C., 2000, astro-ph/0011437

\bibitem[1997]{Pettini1}
Pettini M., Smith L.\,J., King D.\,L., Hunstead R.\,W., 1997, ApJ 486, 665 (P97)

\bibitem[]{Pettini2}
Pettini M., Kellogg M., Steidel C.\,C., Dickinson M., Adelberger~K.\,L., Giavalisco M., 1998, ApJ 508, 539 

\bibitem[]{Pfenniger}
Pfenniger, D., Combes, F., Martinet, L. 1994 A\&A 285, 79

\bibitem[]{Pozzetti}
Pozzetti, L., Madau, P., Ferguson, H.C., Zamorani, G., \& Bruzual, G.A, 1998, MNRAS 298, 1133

\bibitem[]{Prantzos}
Prantzos, N. and Boissier, S., 2000, MNRAS 315, 82

\bibitem[]{PABB}
Puget J.-L., Abergel A., Bernard J.-P., Boulanger F., Burton~W.\,B., D\'esert
F.-X., Hartmann D., 1996, A\&A 308, L5


\bibitem[1981]{Renzini}
Renzini, A. and Voli, M., 1981, A\&A 94, 175

\bibitem[1997]{Renzini}
Renzini, A., 1997, ApJ 488, 35

\bibitem[1997]{Sanders}
Sanders, D.B. \& Mirabel, I.F, 1996, ARA\&A, 34, 749

\bibitem[1997]{Savaglio}
Savaglio, S., 2000, in the Proceedings of IAU symposium 204 "The extragalactic infrared background and its implications on cosmology implications" Manchester, August 2000, Eds M. Hartwitt, \& M.G. Hauser 

\bibitem[1993a]{Schaerer}
Schaerer, D., Charbonnel, C., Meynet, G., Maeder, A. and Schaller, G., 1993a, A\&AS 102, 339

\bibitem[1993b]{Schaerer}
Schaerer, D., Meynet, G., Maeder, A. and Schaller, G., 1993b, A\&AS 98, 523

\bibitem[1992]{Schaller}
Schaller, G., Schaerer, D., Meynet, G. and Maeder, A., 1992, A\&AS 96, 269

\bibitem[1998]{Silva}
Silva, L., Grabnato, G.L., Bressan, A. and Danese, L., 1998, ApJ 509, 103

\bibitem[1991]{Soifer} 
Soifer, B.~T., and Neugebauer, G. 1991, AJ 101, 354

\bibitem[1997]{} 
Songaila, A., 1997, ApJ 490, L1

\bibitem[1996]{} 
Songaila, A. \& Cowie, L.L., 1996, AJ 112, 335

\bibitem[]{}
Smail, I, Ivison, R.J., Blain, A.W., 1997, ApJ 490, 5

\bibitem[]{}
Steidel, C.C, Adelberger, K.L, Giavalisco, M, Dickinson, M., and Pettini, P., 1999, ApJ 519, 1


\bibitem[1996]{Storrie}
Storrie-Lombardi, L., McMahon, R., and Irwin, M. 1996, MNRAS 283, L79

\bibitem[1997]{Treyer}
Treyer, M.A., Ellis, R.S., Milliard, B., and Donas, J., 1997 in 
The Ultraviolet Universe at Low and High Redshift, ed. W. Waller,
(Woodbury: AIP Press), in press

\bibitem[1980]{Tinsley}
Tinsley, B.M., 1980., Fund. Cosmic Phys., 5, 287

\bibitem[1995]{Tytler}
Tytler, D. {\it et al.} 1995, in QSO Absorption Lines, ESO Astrophysics Symposia, edited by G. Meylan (Springer, Heidelberg), p. 289

\bibitem[1995]{Viegas}
Viegas, SueliM. 1995, MNRAS 276, 268

\bibitem[1995]{Vogel}
Vogel, S., Weymann, R., Rauch, M, Hamilton, T., 1995, ApJ 441, 162


\bibitem[1991]{Young}
Young, J.S. and Scoville, N.Z. 1991, ARA\&A 29, 581

\bibitem[2000]{Zoccali}
Zoccali, M., et al., 2000, ApJ 530, 418
\end{thebibliography}
\end{document}